\documentclass[12pt,letterpaper]{article}
\usepackage{graphicx}
\usepackage{amsmath}

\input{tcilatex}

\begin{document}

\author{V. V. Prosentsov\thanks{%
e-mail: prosentsov@yahoo.com} \\
Stationsstraat 86, 5751 HH, Deurne, The Netherlands}
\title{Scattering optimization of photonic cluster: from minimal to maximal
reflectivity}
\maketitle

\begin{abstract}
The optimization of the light scattered by photonic cluster made of small
particles is studied with the help of the local perturbation method and
special optimization algorithm. It was shown that photonic cluster can be
optimized in a such a way that its reflectivity will be increased or
decreased by several orders of magnitude for selected wavelength and
direction.
\end{abstract}

\section{Introduction}

Light is known as fastest carrier of energy and information and this
property makes light indispensable for communications, warfare and
fundamental research. While light can be relatively easy created and guided,
its manipulation is somewhat difficult. Manipulation of the light requires
dynamical control of the refractive index of the host medium. While the
materials for active control of the light are still under development, the
theoretical studies are already started. Recently, the light manipulation
was investigated in work \cite{Aperiodic} where the weak scattering by the
finite object was studied by using the first Born approximation. In
practice, the light scattering by photonic cluster may be not weak and it
should be studied with other methods. The local perturbation method (LPM) is
suitable tool for such analysis. The LPM correctly describes scattering by
particles with arbitrary large refractive index which are small compared to
the incident wavelength and the scattering can be strong (see for example
work \cite{Chaumet}-\cite{BassVpFr} and references wherein). In works \cite
{VpAd1}-\cite{Vp} the LPM was used to study the wave propagation in the
photonic cluster.

In this paper we optimize the light scattering from the cluster made of
small particles by using the LPM and special optimization technic. By using
our method we modify the cluster in such a way that the scattering from the
cluster is significantly minimized or maximized at one point. We present
several examples which will show that the scattering by the cluster can be
increased and decreased by several orders of magnitude for selected
wavelength and direction.

\section{The formalism}

The formalism we use is presented in many works (see for example \cite
{Chaumet}-\cite{BassVpFr}) and we only briefly present it here for
convenience and consistency. Consider the photonic cluster made of particles
which characteristic sizes are small compared to the incident wavelength $%
\lambda _{0}$. The electric field $\mathbf{E}$\ propagating in the host
medium filled with $N$ particles is described by the following equation \cite
{BassVpFr}

\begin{equation}
\left( \bigtriangleup -\mathbf{\nabla }\otimes \mathbf{\nabla }%
+k_{0}^{2}\right) \mathbf{E}(\mathbf{r})+\frac{\omega ^{2}}{c^{2}}%
\sum_{n=0}^{N-1}(\varepsilon _{sc,n}-\varepsilon _{0})f_{n}(\mathbf{r}-%
\mathbf{r}_{n})\mathbf{E}(\mathbf{r}_{n})=\mathbf{S}(\mathbf{r}),
\label{inv11}
\end{equation}
where

\begin{equation*}
\mathbf{\;}k_{0}=\left| \mathbf{k}_{0}\right| =\frac{2\pi }{\lambda _{0}}=%
\frac{\omega }{c}\sqrt{\varepsilon _{0}},\;f_{n}(\mathbf{r}-\mathbf{r}%
_{n})=\left\{ 
\begin{array}{cc}
1\text{,} & \text{inside particle } \\ 
0\text{,} & \text{outside particle}
\end{array}
\right. .
\end{equation*}
Here $\bigtriangleup $ is Laplacian and $\mathbf{\nabla }$\ nabla operators, 
$\otimes $ defines tensor product, $k_{0}$ is a wave number in the host
medium ($\omega $ is the angular frequency and $c$ is the light velocity in
vacuum), $\varepsilon _{sc,n}$ and $\varepsilon _{0}$ are the permittivity
of the $n$-th particle and the medium respectively, $f_{n}$ is the function
describing the shape of the $n$-th scatterer, and $\mathbf{S}$ is the field
source. The characteristic size of $n$-th scatterer we denote as $L_{n}$.
Note, that the equation (\ref{inv11}) is an approximate one and this is true
only when the small scatterers ($k_{0}L_{n}\ll 1$) are considered. This
equation is easily solvable with respect to the fields $E(\mathbf{r}_{n})$
when the positions $\mathbf{r}_{n}$ are known (it will invoke solution of $%
3N $ linear equations for each frequency $\omega )$. The solution of the
equation (\ref{inv11}) can be written in the following form

\begin{equation}
\mathbf{E}(\mathbf{r})=\mathbf{E}_{0}(\mathbf{r})+\mathbf{E}_{sc}(\mathbf{r}%
),  \label{inv12}
\end{equation}
where

\begin{equation}
\mathbf{E}_{sc}(\mathbf{r})=\frac{\omega ^{2}}{c^{2}}\left( \widehat{I}+%
\frac{\mathbf{\nabla }\otimes \mathbf{\nabla }}{k_{0}^{2}}\right)
\sum_{n=0}^{N-1}\mathbf{E}(\mathbf{r}_{n})(\varepsilon _{sc,n}-\varepsilon
_{0})\Phi _{n}(\mathbf{r})  \label{inv13}
\end{equation}
and

\begin{equation}
\Phi _{n}(\mathbf{r})=\int_{-\infty }^{\infty }\frac{\widetilde{f_{n}}(%
\mathbf{q})e^{i\mathbf{q\cdot (r-r}_{n})}}{(q^{2}-k_{0}^{2})}d\mathbf{q,\;}\;%
\widetilde{f_{n}}(\mathbf{q})=\frac{1}{8\pi ^{3}}\int_{-\infty }^{\infty
}f_{n}(\mathbf{r})e^{-i\mathbf{q\cdot r}}d\mathbf{r.}  \label{inv14}
\end{equation}
Here $\widehat{I}$ is the $3\times 3$ unitary tensor in polarization space
and $\mathbf{r}_{n}$ is the radius vector of the n-th particle. The field $%
\mathbf{E}(\mathbf{r}_{n})$ is the field inside the $n$-th particle, $%
\widetilde{f_{n}}$ is the Fourier transform of the function $f_{n}$, and $%
\mathbf{\cdot }$ defines scalar product. The incident field $\mathbf{E}_{0}$
is created by the source $\mathbf{S}$ in the host medium and it is not
important for our consideration.

The formula (\ref{inv12}) is rather general one and it describes the field
in the medium with photonic cluster of arbitrary form made of small
particles of arbitrary form.

The fields $\mathbf{E}(\mathbf{r}_{n})$ should be found by solving the
system of $3N$ linear equations obtained by substituting $\mathbf{r}=\mathbf{%
r}_{n}$ into Eq. (\ref{inv12}). The formula for the scattered field (\ref
{inv13}) can be simplified even further when the distance between the
observer and an $n$-th scatterer ($R_{n}$) is large, i.e. when $R_{n}\gg
L_{n}$. In this case the integral $\Phi _{n}$ can be calculated
approximately. We note also that the integral $\Phi _{n}$ can be calculated
exactly at least for the spherical particles. When $R_{n}\gg L_{n}$
integration in (\ref{inv13}) gives

\begin{equation}
\mathbf{E}_{sc}(\mathbf{r})=\frac{\omega ^{2}}{4\pi c^{2}}\left( \widehat{I}+%
\frac{\mathbf{\nabla }\otimes \mathbf{\nabla }}{k_{0}^{2}}\right)
\sum_{n=0}^{N-1}\mathbf{E}(\mathbf{r}_{n})(\varepsilon _{sc,n}-\varepsilon
_{0})\frac{e^{ik_{0}R_{n}}}{R_{n}}V_{n},  \label{inv15g}
\end{equation}
where

\begin{equation}
R_{n}=\left| \mathbf{r}-\mathbf{r}_{n}\right| \gg L_{n}.  \label{inv16}
\end{equation}
Here $R_{n}$ is the distance between the observation point and $n$-th
scatterer, $V_{n}$ is the volume of the $n$-th scatterer, and $\left| 
\mathbf{...}\right| $ brackets denote absolute value.

In practice, the distance between the cluster and the observer is much more
larger than the size of the cluster and the inequality $\left| \mathbf{r}%
\right| \gg \max (\left| \mathbf{r}_{n}\right| )$ is fulfilled. In this case
the field (\ref{inv15g}) can be simplified and rewritten in the form

\begin{equation}
\mathbf{E}_{sc}(\mathbf{r})=\frac{\omega ^{2}e^{ik_{0}r}}{4\pi c^{2}r}\left( 
\widehat{I}-\mathbf{l}\otimes \mathbf{l}\right) \sum_{n=0}^{N-1}\mathbf{E}(%
\mathbf{r}_{n})V_{n}(\varepsilon _{sc,n}-\varepsilon _{0})e^{-ik_{0}\mathbf{%
l\cdot r}_{n}},  \label{inv16c}
\end{equation}
where

\begin{equation*}
\mathbf{l}=\mathbf{r}/r,\;r=\left| \mathbf{r}\right| \gg \max (\left| 
\mathbf{r}_{n}\right| ).
\end{equation*}

\section{Optimization of the cluster reflectivity}

The idea of our approach is the following. We assume that scatterers in the
cluster can be repositioned in a such a way that the field scattered by the
cluster will be tuned to required value at the observation point.

At first, we calculate the intensity of the scattered field from the given
(not modified) cluster for selected wavelength $\lambda _{opt}$ and
observation point $\mathbf{r}$. After this we select one particle in the
cluster and reposition it a bit in $x$, $y$, and $z$ directions (usually for
tenth of the particle size) recalculating the scattered field in $\mathbf{r}$
each time we move the particle. From the array of the calculated intensities
we select the value nearest to the required one and we place the particle at
the point corresponding to this intensity value. We perform this procedure
for all particles in the cluster and repeat it several times. After several
iterations, the particles in the cluster will be rearranged in such a way
that the intensity of the scattered light will correspond the required value
at the selected observation point $\mathbf{r}$\ and for the wavelength $%
\lambda _{opt}$.

\begin{figure}[t]
\begin{center}
\includegraphics [width=9cm]
{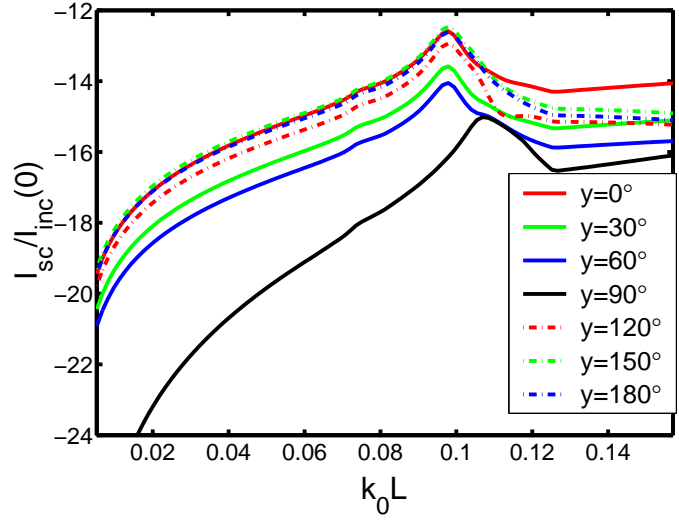}
\end{center}
\caption{The logarithm of the normalized intensity of the scattered field $%
I_{sc}(\mathbf{r})/I_{inc}(0)$ versus $k_{0}L$ for the not optimized $Au$
cluster. The particles in the cluster are arranged into cubic lattice. The
permittivity of the host medium is $\protect\varepsilon _{0}=1$, the
characteristic size of the particles is $L=10$ nm, the period of the cluster
is $d=3.3L$, and the total number of the particles is $N=123$. The source is
positioned at $\mathbf{r}_{s}=\{x_{s},0,0\}$\ and the observation points $%
\mathbf{r}$ are located at the circle with radius $x_{s}$. The angle $%
\protect\theta $ is the angle between vector $\mathbf{r}$ and $\mathbf{r}%
_{s} $ in $xy$ plane.}
\label{fig1}
\end{figure}

For example, super reflective cluster can be created when the intensity of
the scattered field $I_{sc}=\left| \mathbf{E}_{sc}(\mathbf{r})\right| ^{2}$
will be maximized at some point and minimized at all other points
(directions). Another important example closely related to optical cloaking
is the cluster with significantly reduced reflectivity for selected
wavelength at chosen direction. When the reflectivity of the cluster is
lower than the sensitivity level of the receiver, this cluster is actually
invisible for observer.

Moreover, when the intensity of the field scattered by the cluster is
changed in some direction, this can be used in directional optical switch.
This kind of device can be used for light houses, for example. When the
spectrum of the field scattered by the cluster is modified in time when the
cluster can be used as a filter.

Below we present several examples demonstrating the possibility to optimize
the cluster reflectivity in chosen direction for selected wavelength.

We note that similar approach was successfully used in work \cite{Ivo} where
the phase of incident field was tuned to maximize the field scattered from
the complex object (eggshell and TiO powder) in chosen direction.

\begin{figure}[t]
\begin{center}
\includegraphics [width=9cm]
{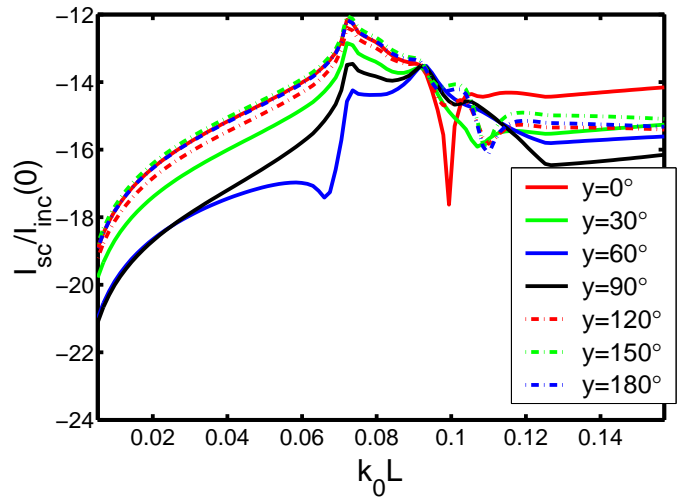}
\end{center}
\caption{The logarithm of the normalized intensity of the scattered field $%
I_{sc}(\mathbf{r})/I_{inc}(0)$ versus $k_{0}L$ for the optimized spherical
cluster made of $Au$ cubes. The permittivity of the host medium is $\protect%
\varepsilon _{0}=1$, the characteristic size of the cubes is $L=10$ nm and
the total number of the particles is $N=123$. The source is positioned at $%
\mathbf{r}_{s}=\{x_{s},0,0\}$\ and the observation points $\mathbf{r}$ are
located at the circle with radius $x_{s}$. The angle $\protect\theta $ is
the angle between vector $\mathbf{r}$ and $\mathbf{r}_{s}$ in $xy$ plane.
The cluster reflectivity was minimized for $k_{0}L=0.1$ at $\mathbf{r}=%
\mathbf{r}_{s}$. The red line shows extremely low deep in cluster
reflectivity after the optimization.}
\label{fig2}
\end{figure}

\subsection{Examples}

Consider the spherical photonic cluster consisting of $Au$ cubes. The
characteristic size of the cubes is $L$ and they are organized into simple
cubic lattice with period $d$. The center of the cluster is positioned at
the beginning of the coordinates. The incident field is generated by the
point source positioned at the point $\mathbf{r}_{s}=\{x_{s},0,0\}$ and it
is linearly polarized in $y$ direction. The scattering by the cluster will
be optimized (to minimum or maximum) for the chosen wavelength $\lambda
_{opt}$ for the\ observer positioned at the point $\mathbf{r}$. We assume
that the permittivity of the scatterers is the same as the permittivity of
the bulk $Au$ and the actual values of permittivity were taken from \cite
{Palik}. Note that the source and the observer are far from the cluster such
that $k_{0}\left| \mathbf{r}_{s}\right| \gg 1$ and $k_{0}\left| \mathbf{r}%
\right| \gg 1$.

\begin{figure}[t]
\begin{center}
\includegraphics [width=9cm]
{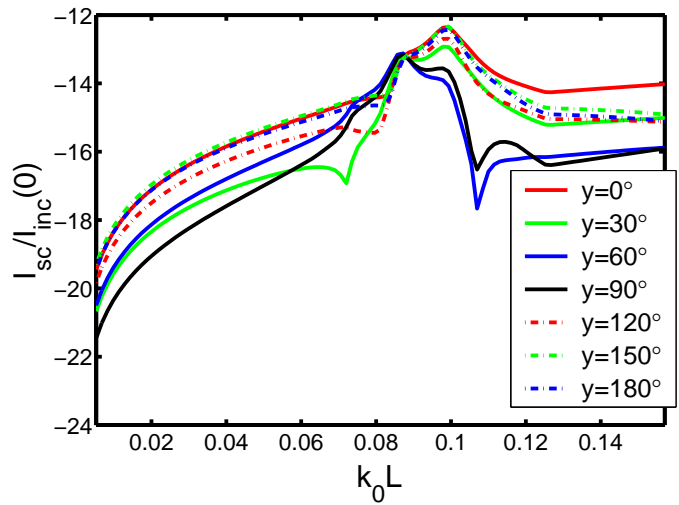}
\end{center}
\caption{The logarithm of the normalized intensity of the scattered field $%
I_{sc}(\mathbf{r})/I_{inc}(0)$ versus $k_{0}L$ for the optimized spherical
cluster made of $Au$ cubes. The permittivity of the host medium is $\protect%
\varepsilon _{0}=1$, the characteristic size of the cubes is $L=10$ nm and
the total number of the particles is $N=123$. The source is positioned at $%
\mathbf{r}_{s}=\{x_{s},0,0\}$\ and the observation points $\mathbf{r}$\ are
located at the circle with radius $x_{s}$. The angle $\protect\theta $ is
the angle between vector $\mathbf{r}$ and $\mathbf{r}_{s}$ in $xy$ plane.
The cluster reflectivity was maximized for $k_{0}L=0.1$ at $\mathbf{r}=%
\mathbf{r}_{s}$. The red line shows extremely high peak in cluster
reflectivity after the optimization.}
\label{fig3}
\end{figure}

\begin{figure}[t]
\begin{center}
\includegraphics [width=9cm]
{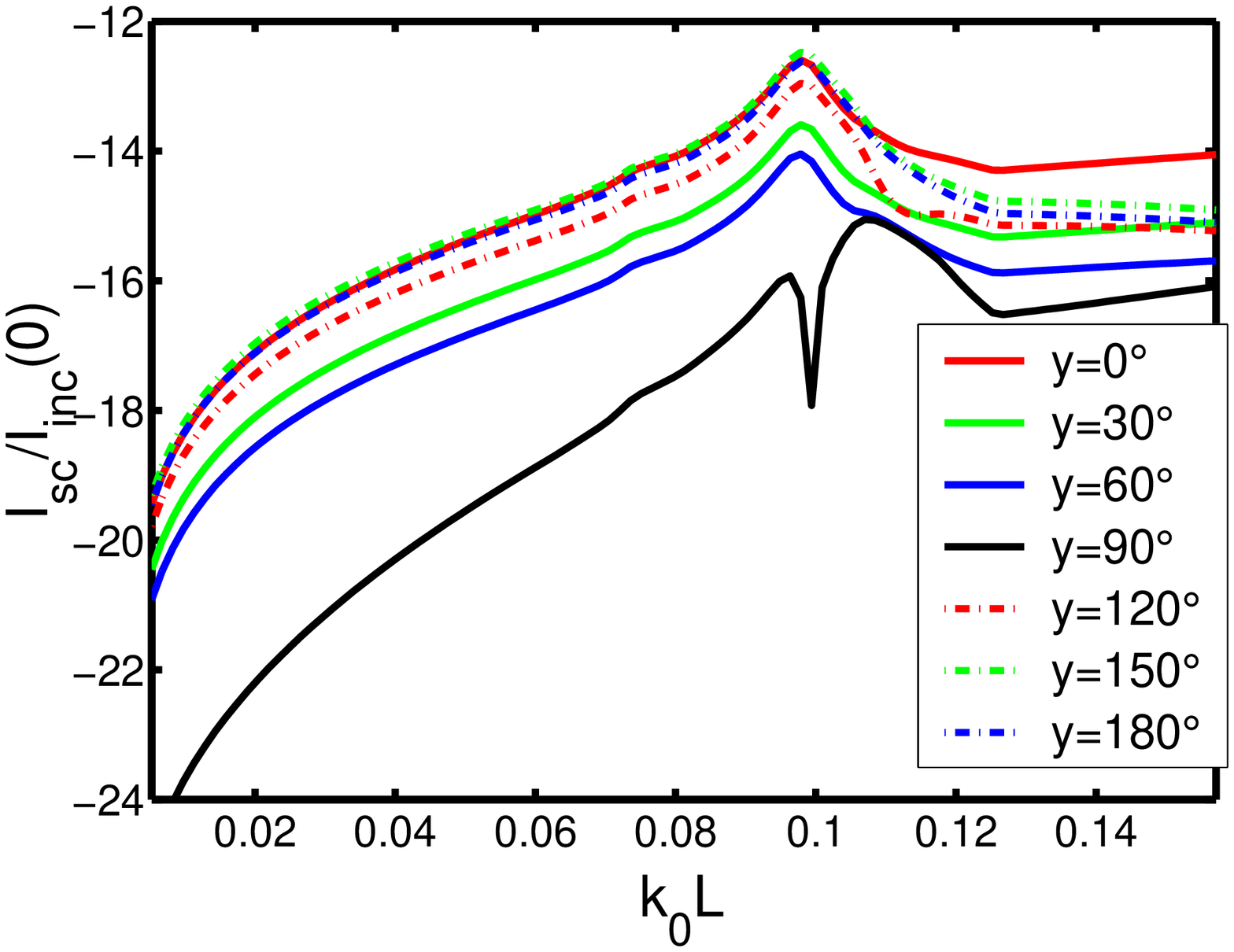}
\end{center}
\caption{The logarithm of the normalized intensity of the scattered field $%
I_{sc}(\mathbf{r})/I_{inc}(0)$ versus $k_{0}L$ for the optimized spherical
cluster made of $Au$ cubes. The permittivity of the host medium is $\protect%
\varepsilon _{0}=1$, the characteristic size of the cubes is $L=10$ nm and
the total number of the particles is $N=123$. The source is positioned at $%
\mathbf{r}_{s}=\{x_{s},0,0\}$\ and the observation points $\mathbf{r}$ are
located at the circle with radius $x_{s}$. The angle $\protect\theta $ is
the angle between vector $\mathbf{r}$ and $\mathbf{r}_{s}$ in $xy$ plane.
The cluster reflectivity was minimized for $k_{0}L=0.1$ at $\mathbf{r}%
=\{0,x_{s},0\}$. The red line shows extremely low deep in cluster
reflectivity after the optimization.}
\label{fig4}
\end{figure}

\begin{figure}[t]
\begin{center}
\includegraphics [width=9cm]
{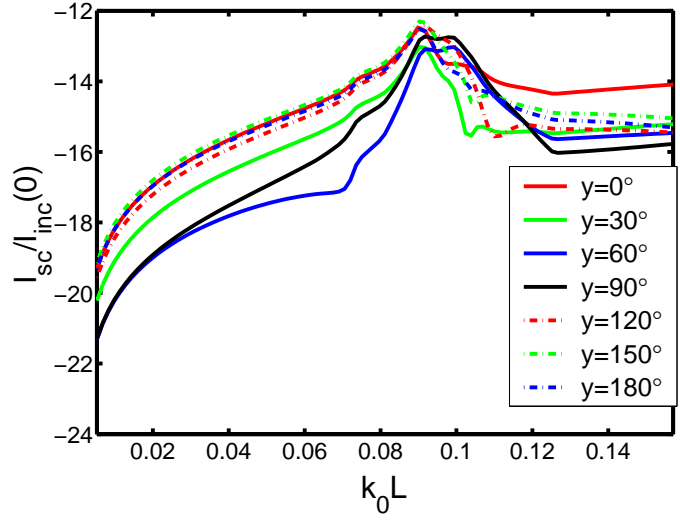}
\end{center}
\caption{The logarithm of the normalized intensity of the scattered field $%
I_{sc}(\mathbf{r})/I_{inc}(0)$ versus $k_{0}L$ for the optimized spherical
cluster made of $Au$ cubes. The permittivity of the host medium is $\protect%
\varepsilon _{0}=1$, the characteristic size of the cubes is $L=10$ nm and
the total number of the particles is $N=123$. The source is positioned at $%
\mathbf{r}_{s}=\{x_{s},0,0\}$\ and the observation points $\mathbf{r}$ are
located at the circle with radius $x_{s}$. The angle $\protect\theta $ is
the angle between vector $\mathbf{r}$ and $\mathbf{r}_{s}$ in $xy$ plane.
The cluster reflectivity was maximized for $k_{0}L=0.1$ at $\mathbf{r}%
=\{0,x_{s},0\}$. The black line shows extremely high peak in cluster
reflectivity after the optimization.}
\label{fig5}
\end{figure}

The examples of the optimization are presented on Figs. \ref{fig1}-\ref{fig5}%
. The figures show the normalized intensity of the light scattered by the
clusters versus $k_{0}L$. The Fig. \ref{fig1} shows the light intensity
scattered from the not optimized cluster. Figs. \ref{fig2} and \ref{fig3}
show the results for the clusters which scattering was optimized at the
point $\mathbf{r}=\mathbf{r}_{s}$ and the Figs. \ref{fig4} and \ref{fig5}
show the results for the clusters which scattering was optimized at the
point $\mathbf{r}=\{0,x_{s},0\}$.

The positions of the particles in the clusters were changed in order to
maximize or minimize the scattering for the wavelength $\lambda _{opt}=633$
nm ($k_{0}L=0.1$) in chosen directions. The results are presented for
different angles\ of observation $\theta $ in $xy$ plane. The Fig. \ref{fig2}
shows normalized (with respect to the intensity of the incident field $%
I_{inc}(0)$) intensity of the light $I_{sc}(\mathbf{r})=\left| \mathbf{E}%
_{sc}(\mathbf{r})\right| ^{2}$ scattered by the optimized photonic cluster.
The cluster was optimized to minimal intensity of the scattered field in the
direction of the source ($\theta =0$). Comparing Fig. \ref{fig1} and Fig. 
\ref{fig2} one can see huge deep (about five orders of magnitude) in the
intensity of the scattered field for the selected wavelength $\lambda _{opt}$
and the selected direction $\theta =0$. The Fig. \ref{fig2} shows also that
unintendently, the optimization significantly increased intensity of the
scattered field for other wavelengths (near $k_{0}L=0.07$) in almost all
directions.

The Fig. \ref{fig3} shows normalized intensity of the light scattered by the
cluster optimized to maximal intensity of the scattered field in the
direction $\theta =0$. Comparing Fig. \ref{fig1} and Fig. \ref{fig3} one can
see relatively large increase (about two times) in the intensity of the
scattered field for the selected wavelength $\lambda _{opt}$ and the
selected direction $\theta =0$. The Fig. \ref{fig3} shows also that the
optimization significantly increased intensity of the scattered field for $%
k_{0}L=0.09$ and significantly decreased the intensity for $k_{0}L=0.11$ in
directions with $\theta =\pi /2$ and $\theta =\pi /3$ respectively.

The Fig. \ref{fig4} shows normalized intensity of the light scattered by the
cluster optimized to minimal intensity of the scattered field in the
direction perpendicular to the source-cluster direction ($\theta =\pi /2$).
Comparing Fig. \ref{fig1} and Fig. \ref{fig4} one can see significant
decrease (about two orders of magnitude) in the intensity of the scattered
field for the selected wavelength $\lambda _{opt}$ in the selected direction 
$\theta =\pi /2$. In distinction to previous examples, the intensity of the
scattered field in other directions is almost not affected in this case.

The Fig. \ref{fig5} shows normalized intensity of the light scattered by the
cluster optimized to maximal intensity of the scattered field in the
direction $\theta =\pi /2$. Comparing Fig. \ref{fig1} and Fig. \ref{fig5}
one can see huge increase (about three orders of magnitude) in the intensity
of the scattered field for the selected wavelength $\lambda _{opt}$ in the
selected direction $\theta =\pi /2$. We note that the intensity of the
scattered field increased also in direction $\theta =\pi /3$ (for $\lambda
_{opt}$) and for all other directions the intensity peak shifted to $%
k_{0}L=0.09$.

Examination of the Figs. \ref{fig1}-\ref{fig5} reveals several important
things. The first one is that the optimization works because the intensity
of the light scattered by the optimized cluster has very clear minima or
maxima for the chosen wavelengths and positions. The second one is that the
intensity of the scattered field has deep or peak not only for designed
wavelength but for other wavelengths and directions. It should be emphasized
that the scattering was optimized for one direction and for one wavelength
only and that is why the optimization is effective for narrow spectrum of
wavelengths and scattering directions.

\section{Conclusions}

The reflectivity of the spherical photonic cluster made of small particles
was optimized to maximal and minimal values by using the local perturbation
method and the special optimization algorithm. The possibility to design the
photonic cluster with required scattering characteristics was demonstrated
for selected wavelength and direction.

\begin{equation*}
\end{equation*}
\textbf{Acknowledgments}

I would like to express my gratitude to Prof. Valentin Freilikher for
critical comments and helpful discussions. I would like to thank also Ivan
Nikolaev for his comments and important suggestions.

\begin{equation*}
\end{equation*}

\end{document}